\documentclass{article}
\usepackage{amssymb}

\usepackage{graphicx}
\usepackage{amsmath}

\input{tcilatex}

\begin{document}

\title{Shubnikov-de Haas effect in quasi-two-dimensional compounds}
\author{P.D. Grigoriev$^{1,2}$, M.V. Kartsovnik$^{3}$, W. Biberacher$^{3}$and P.
Wyder$^{1}$ \\
$^{1}$Grenoble High Magnetic Field Laboratory, MPI-FKF and CNRS, \\
BP 166, F-38042 Grenoble Cedex 09, France \\
$^{2}$L.D.Landau Institute for Theoretical Physics, 142432 Chernogolovka,
Russia \\
$^3$ Walther-Mei\ss ner-Institut, Bayerische Akademie der Wissenschaften, \\
Walther-Mei\ss ner-Str. 8, D-85748 Garching, Germany}
\date{\today }
\maketitle

\begin{abstract}
The Shubnikov - de Haas effect in quasi-two-dimensional compounds is
studied. The conductivity is calculated from the Kubo formula. Two effects
-- the field-dependent phase shift of the beats and the slow oscillations of
the conductivity are explained and calculated. The results are applicable to
the strongly anisotropic organic metals and other layered compounds.
\end{abstract}

\section{Introduction}

The magnetic quantum oscillations (the de Haas - van Alphen (dHvA) and the
Shubnikov - de Haas (SdH) effects) have been discovered long ago and were
used a lot as a powerful tool of studying the geometry of Fermi surfaces and
other electronic properties of different metals \cite{Sh}. Last years the
quasi-two-dimensional (Q2D) organic metals \cite{OMRev} attract a great
interest and many works have been devoted to the study of magnetic quantum
oscillations in these compounds (for a review see e.g. \cite{MQORev}). The
quantum oscillations of the magnetization is a thermodynamic effect that is
completely determined by the density of states distribution. An exact
calculation of the density of states is a very complicated problem but the
semi-phenomenological theoretical description of the magnetization
oscillations in Q2D compounds was recently provided in a number of papers 
\cite{Gv1,Vagner,Harrison,IMV,GV,Champel,Pavel2}. The chemical potential
oscillations and the arbitrary electron reservoir due to the open sheets of
the Fermi surface make no principal difficulties\cite{Pavel2}. Of course, if
the e-e interaction drastically changes the ground state (like the FQHE
effect) the problem becomes much more complicated. But when the number of
the occupied LLs is very large ($n_{F}>100$ as in most of Q2D organic
metals) the effect of the e-e interaction is reduced (as in the Fermi
liquid) and can be taken into account via the renormalization of the
effective mass. Another open question in the theory of the Q2D dHvA effect
is the exact shape of the Landau levels. It depends on the particular type
of a compound (on the type of impurities, their distribution, the interlayer
transfer integral etc.) and makes only some quantitative differences.

The theoretical description of the Q2D quantum conductivity oscillations was
not as successful although some works on this subject have appeared last
years \cite{Harrison,DS,Mc,Gv2,PhSh}. There are still some open qualitative
questions. One of them is the origin of the phase difference in the beats of
the SdH and dHvA effects. It takes place even when the harmonic damping is
very strong and only the first harmonic is seen. An explanation and an
approximate theoretical description of this phenomenon as well as the more
extensive experimental study of this effect has been proposed recently \cite
{PhSh}.

Another very interesting phenomenon is the slow oscillations of the
magnetoresistance that have been observed in a number of Q2D organic metals 
\cite{MQORev,ibr2,i3,tera,ohmi,togo,broo}. The behavior of the slow
oscillations resembles that of the SdH effect that has lead to a suggestion
of additional, very small Fermi surface pockets in these materials. However,
band structure calculations (basically giving a good description of the
electron band structure and Fermi surface topology in the organic metals)
have shown no evidence of such small pockets in any of these compounds.
Moreover, while the slow oscillations are often very pronounced in the
magnetoresistance, sometimes even dominating the oscillation spectrum, no
analogous observation in oscillating magnetization (dHvA effect) has been
reported thus far. Certainly, the slow oscillations must contain a useful
information about the compounds but to extract this information one needs
some theoretical explanation and, desirably, a quantitative description of
the phenomenon.

In this paper we calculate the conductivity starting from the Kubo formula
and handling carefully with all oscillating quantities in the expression of
the conductivity. The proposed theory explains both the slow oscillations
and the phase shift of the beats and gives some quantitative description of
these phenomena. It should be noted, however, that our analysis has one
important limitation -- it is developed only when the magnetic field is
perpendicular to the conducting layers. When the magnetic field has
substantial tilting angle our results give only a qualitative description.

In section II we calculate the conductivity starting from the Kubo formula.
The calculations are described in detail so that they can be easily
followed. In Sections III and IV we study the limit of strong harmonic
damping in more details. In Sec.\ III we give the description of the phase
shift of beats. In Sec. IV the slow oscillations are considered.

\section{The calculation of the conductivity}

First, we shall consider the SdH effect in the external magnetic field
perpendicular to the conducting planes: $\vec{H}\Vert \vec{z}$. Then the
electron spectrum of Q2D electron gas in magnetic field is given by 
\begin{equation}
\epsilon _{n,k_{z}}=\hbar \omega _{c}\,(n+\frac{1}{2})-2t\cos (k_{z}d)
\label{ES}
\end{equation}
where $\omega _{c}=eB_{z}/m^{\ast }c$ is the cyclotron frequency, $k_{z}$ is
the wave number in $z$-direction perpendicular to the conducting layers, $d$
is the interlayer distance and $t$ is the interlayer transfer integral. This
energy spectrum can be easily obtained in the tight binding approximation
after we choose the electromagnetic vector potential in the form $%
A=(0,B_{z}x,0)$. The interlayer bandwidth $W$ is related to the interlayer
transfer integral as $W=4t.$ We consider the case $\epsilon _{F}\gg \hbar
\omega _{c}$. Then the electron-electron interaction does not completely
change the energy spectrum (as in the extremely quantum limit of FQHE
regime) and we can take the energy spectrum (\ref{ES}) as a zeroth
approximation and then treat the impurity scattering using the perturbation
theory. If magnetic field is tilted with respect to the conducting planes
the formula (\ref{ES}) becomes no more valid. This is the most important
limitation of the analysis that follows. Hence, at the substantial tilting
angles our theory gives only qualitative predictions.

Before diving into the calculations let us describe on the physical level
the origin of the slow oscillations. They come from the interference of the
oscillations of the different quantities, on which the conductivity depends.
These quantities are the electron relaxation time, the mean square electron
velocity (summed over all states on the FS) and the Dingle temperature. The
product of two oscillating factors gives a constant term that is just one
half of the product of the amplitudes of these oscillations: $(1+a\cos
x)(1+b\cos x)=1+(a+b)\cos x+(ab/2)\cos 2x+ab/2$ (where in our case $x\equiv 
\frac{2\pi \mu }{\hbar \omega _{c}}$). The last term $ab/2$ is responsible
for the slow oscillations. In our case the amplitudes $a,b$ of the fast
oscillations (for example, of relaxation time and the mean square velocity)
are themselves the slowly oscillating functions of the magnetic field due to
the beating. Hence, the conductivity contains the terms $\sim R_{D}^{2}\cos
\left( 4\pi t/\hbar \omega _{c}\right) ^{2}=(R_{D}^{2}/2)(1+\cos \left( 8\pi
t/\hbar \omega _{c}\right) )$. The term $(R_{D}^{2}/2)\cos \left( 8\pi
t/\hbar \omega _{c}\right) $ describes the slow oscillations with the
frequency equal to the double beating frequency. The amplitude of the slow
oscillations contains the square of the Dingle factor. Nevertheless, it can
be larger than the amplitude of the fast SdH oscillations. The slow
oscillations depend only on the transfer integral $t$ and are not sensitive
to the exact position of the chemical potential. Hence, the amplitude of the
slow oscillations does not contain the temperature smearing factor and is
larger than the amplitude of the Shubnikov oscillations at high enough
temperature (compared to the Dingle temperature).

The magnetization does not have such a slow oscillating term because in the
lowest order on damping factors [see Eq. (\ref{M}) below; the term $%
\,(1+b\cos x)$ comes, for example, from the Dingle factor] it has only the
interference of $a\sin x\,(1+b\cos x)=1+a\sin x+(ab/2)\sin 2x$ that does not
have a constant term $ab/2$. Hence in the same lowest order on the damping
factors the magnetization does not reveal the slow oscillations.

The origin of the phase shift was explained recently in \cite{PhSh} basing
on the Boltzmann transport equation. In this paper we shall give more strict
calculation of this effect.

\bigskip

To calculate the conductivity we shall use the Kubo formula \cite{Mah}. The
procedure is similar to that in three-dimensional metals without magnetic
field (\cite{Mah}, \S\ 7.1.2). In magnetic field only the new set of quantum
numbers $m\equiv \{n,k_{z},k_{y}\}$ instead of the momentum $\vec{p}$ and
the different dispersion relation (\ref{ES}) should be used. The evaluation
of the Kubo formula without vertex corrections gives 
\begin{equation}
\sigma _{zz}=\frac{e^{2}}{V}\sum_{m}v_{z}^{2}(m)\int \frac{d\epsilon }{2\pi }%
A^{2}(m,\epsilon )\left( -n_{F}^{\prime }(\epsilon )\right) \,  \label{Sigd1}
\end{equation}
where the volume $V$ is to normalize the sum over quantum numbers $m$ , $e$
is the electron charge, the limits of the integral over $\epsilon $ are $%
(-\infty ;\infty )$, $\ n_{F}^{\prime }(\epsilon )$ is the derivative of the
Fermi distribution function: 
\begin{equation}
-n_{F}^{\prime }(\epsilon )=1/\{4T\cosh ^{2}\left[ (\epsilon -\mu )/2T\right]
\}  \label{nFd}
\end{equation}
and $A(m,\epsilon )$ is the spectral function that is related to the
electron Green's function $G^{R}(m,\epsilon )$ or to the retarded self
energy part $\Sigma ^{R}(m,\epsilon ):$ 
\begin{equation}
A(m,\epsilon )\equiv -2\mbox{Im}G^{R}(m,\epsilon )=\frac{-2\mbox{Im}\Sigma
^{R}(m,\epsilon )}{\left[ \epsilon -\epsilon (m)-\mbox{Re}\Sigma
^{R}(m,\epsilon )\right] ^{2}+\left[ \mbox{Im}\Sigma ^{R}(m,\epsilon )\right]
^{2}}  \label{SFD}
\end{equation}
The formula (\ref{Sigd1}) is close to the corresponding formula without
magnetic field [\cite{Mah}, formula\ 7.1.10] until the self energy part $%
\Sigma ^{R}(m,\epsilon )$ is specified. It arises mainly from the impurity
scattering. The constant part of the real part $\mbox{Re}\Sigma
^{R}(m,\epsilon )$ of the self energy does not influence the physical
effects. It produces only a constant shift of the chemical potential. The
small oscillations of $\mbox{Re}\Sigma ^{R}(m,\epsilon )$ (that are $\ll
\hbar \omega _{c}$ can also be neglected because the final answer is not
sensitive to $\mbox{Re}\Sigma ^{R}(m,\epsilon )$. Hence, for simplicity, in
the later formulas we shall put $\mbox{Re}\Sigma ^{R}(m,\epsilon )=0$.

The imaginary part of the self energy $\mbox{Im}\Sigma ^{R}(m,\epsilon )$,
on the contrary, is very important since it describes the momentum
relaxation of electrons. The main contribution to the resistivity comes from
the short range impurity scattering. The matrix element of the point-like
impurity scattering is the same for all transitions without change of
electron energy. Hence, the scattering rate $1/\tau (\epsilon )$ is
proportional to the number of states to which an electron may scatter, that
is to the density of states at the given energy. So, if one considers the
point-like impurity scattering in the Born approximation, the imaginary part
of the self energy is proportional to the density of states: 
\begin{equation}
-\mbox{Im}\Sigma ^{R}(m,\epsilon )=C\times \rho (\epsilon )  \label{ImS}
\end{equation}
The constant $C\approx \pi \,C_{i}U^{2}I_{\phi z}$ where $C_{i}$ is the
impurity concentration, $U=\int V(r)d^{3}r$ characterizes the strength of
the impurity potential and $I_{\phi z}$ is a number that depends on the
electron wave function and the impurity distribution along the z-axis. The
physical reason for the proportionality is clear.$1/\tau (\epsilon )\sim
\rho (\epsilon )$. The formula (\ref{ImS}) violates if one considers the
next terms on the small parameter $UN_{LL}/\hbar \omega _{c}=f/d$, where $%
N_{LL}$ is the LL degeneracy per unit area, $f$ -- is the scattering
amplitude (which is constant at small wave vector $q\ll 1/r_{0}$, $r_{0}$ is
the range of the impurity potential) and $d$ is the interlayer distance.
Usually this parameter $f/d$ is really small.

The constant $C$ in (\ref{ImS}) contains many unknown parameters. We shall
not calculate them here but note that $C$ is simply related to the average
Dingle temperature $T_{D}$: 
\begin{equation}
\left\langle \left| \mbox{Im}\Sigma ^{R}(m,\epsilon )\right| \right\rangle
=C\cdot \left\langle \rho (\epsilon )\right\rangle =C\cdot \left(
N_{LL}/\hbar \omega _{c}\right) (1+n_{R})=\pi k_{B}T_{D}  \label{ImSTD}
\end{equation}
where\textbf{\ $\left\langle ..\right\rangle $} means the average value of
something, $k_{B}=1.38\cdot 10^{-16}erg/K$ is the Boltzmann constant and $%
n_{R}$ is the density of reservoir states, that exist in many organic metals
due to the open sheets of the FS.

In the extremely 2D case $(\hbar \omega _{c}\gg t)$ the substantial
deviations from the formula (\ref{ImS}) are possible because the strong
degeneracy of the LLs makes the simple perturbation theory not applicable.
Since we consider the case $2t>\hbar \omega _{c}$ (when the beating of the
oscillations is visible) we shall take\ (\ref{ImS})\ for our subsequent
calculations that are now straightforward.

Performing the summation over $k_{y}$ in (\ref{Sigd1}) and changing the
integration over $k_{z}$\ by the integration over energy $\epsilon (n,k_{z})$
we get 
\begin{eqnarray}
\sigma _{zz} &=&e^{2}N_{LL}\sum_{n}2\int_{0}^{\pi }\frac{d(k_{z}d)}{2\pi }%
v_{z}^{2}(k_{z})\int \frac{d\epsilon }{2\pi }A^{2}(\epsilon
(k_{z},n),\epsilon )\,\left( -n_{F}^{\prime }(\epsilon )\right) =  \notag \\
&=&\,e^{2}N_{LL}d\int 2\frac{d\epsilon ^{\prime }}{2\pi }\sum_{n}\left|
v_{z}(\epsilon ^{\prime },n)\right| \int \frac{d\epsilon }{2\pi }%
A^{2}(\epsilon ^{\prime },\epsilon )\left( -n_{F}^{\prime }(\epsilon )\right)
\label{Sigd2}
\end{eqnarray}
where the velocity $v_{z}(\epsilon ,n)$ is determined from the dispersion
relation (\ref{ES}) 
\begin{eqnarray}
v_{z}(\epsilon ,n) &\equiv &\frac{\partial \epsilon (n,k_{y},k_{z})}{\hbar
\partial k_{z}}=-\frac{2td}{\hbar }\sin (k_{z}d)=  \notag \\
&=&\frac{d}{\hbar }\sqrt{4t^{2}-\left( \epsilon -\hbar \omega
_{c}\,(n+1/2)\right) ^{2}}  \label{vz}
\end{eqnarray}
To go further we have to transform the sum over LLs to the sum over
harmonics. This can be done using the Poisson summation formula (Appendix
A). Substituting (\ref{Sv}) to (\ref{Sigd2}) we get: 
\begin{eqnarray}
\sigma _{zz} &=&e^{2}N_{LL}\int 2\frac{d\epsilon ^{\prime }}{2\pi }%
\sum_{k=-\infty }^{\infty }(-1)^{k}\frac{td^{2}}{\hbar k}\exp \left( \frac{%
2\pi ik\epsilon ^{\prime }}{\hbar \omega _{c}}\right) J_{1}\left( \frac{4\pi
kt}{\hbar \omega _{c}}\right) \int \frac{d\epsilon }{2\pi }A^{2}(\epsilon
^{\prime },\epsilon )\left( -n_{F}^{\prime }(\epsilon )\right) =  \notag \\
&=&e^{2}N_{LL}\sum_{k=-\infty }^{\infty }(-1)^{k}\frac{2td^{2}}{\hbar k}%
J_{1}\left( \frac{4\pi kt}{\hbar \omega _{c}}\right) \int \frac{d\epsilon }{%
2\pi }\left( -n_{F}^{\prime }(\epsilon )\right) I_{z}(\epsilon ,k)
\label{Sigd2*}
\end{eqnarray}
where for the zeroth harmonic $k=0$ one should use the expansion $%
J_{1}(x)=x/2$ at$\,\,\,x\ll 1$, and the integral $I_{z}(\epsilon ,k)$ over $%
\epsilon ^{\prime }$ can be easily evaluated with the spectral function (\ref
{SFD}): 
\begin{eqnarray}
I_{z}(\epsilon ,k) &\equiv &\int \frac{d\epsilon ^{\prime }}{2\pi }%
A^{2}(\epsilon ^{\prime },\epsilon )\exp \left( \frac{2\pi ik\epsilon
^{\prime }}{\hbar \omega _{c}}\right) =  \notag \\
&=&\int \frac{d\epsilon ^{\prime }}{2\pi }\left( \frac{-2\mbox{Im}\Sigma
^{R}(\epsilon )}{\left[ \epsilon -\epsilon ^{\prime }\right] ^{2}+\left[ %
\mbox{Im}\Sigma ^{R}(\epsilon )\right] ^{2}}\right) ^{2}\exp \left( \frac{%
2\pi ik\epsilon ^{\prime }}{\hbar \omega _{c}}\right) =  \notag \\
&=&\exp \left( \frac{2\pi ik\,\epsilon }{\hbar \omega _{c}}\right) \left( 
\frac{1}{\left| \mbox{Im}\Sigma ^{R}(\epsilon )\right| }+\frac{2\pi k}{\hbar
\omega _{c}}\right) R_{D}(k,\epsilon )  \label{Iz}
\end{eqnarray}
where 
\begin{equation}
R_{D}(k,\epsilon )=\exp \left( -2\pi \left| k\right| \,\left| \mbox{Im}%
\Sigma ^{R}(\epsilon )\right| /\hbar \omega _{c}\right)  \label{RD}
\end{equation}
has the form, similar to that of the usual Dingle factor $R_{D}(k)=\exp
\left( -2\pi ^{2}k\,k_{B}T_{D}/\hbar \omega _{c}\right) $\textbf{.}
Collecting the formulas (\ref{Sigd2*} and \ref{Iz}) we get: 
\begin{eqnarray}
\sigma _{zz} &=&e^{2}N_{LL}\int \frac{d\epsilon }{2\pi }\left(
-n_{F}^{\prime }(\epsilon )\right) \sum_{k=-\infty }^{\infty }\frac{%
(-1)^{k}2td^{2}}{\hbar k}J_{1}\left( \frac{4\pi kt}{\hbar \omega _{c}}%
\right) \times  \notag \\
&&\times \exp \left( \frac{2\pi ik\,\epsilon }{\hbar \omega _{c}}\right)
\left( \frac{1}{\left| \mbox{Im}\Sigma ^{R}(\epsilon )\right| }+\frac{2\pi k%
}{\hbar \omega _{c}}\right) R_{D}(k,\epsilon )  \label{Sigs}
\end{eqnarray}

We have to consider all the terms in the expression (\ref{Sigs}) for the
conductivity that make an essential contribution to the oscillations.
Besides the directly oscillating term ($\exp (2\pi ik\epsilon /\hbar \omega
_{c})$ these are the imaginary part of the self energy $\left| \mbox{Im}%
\Sigma ^{R}(\epsilon )\right| $ and the Dingle factor\textbf{\ }$%
R_{D}(k,\epsilon )$. The imaginary part of the self energy is given by (\ref
{ImS}) where the density of states is 
\begin{equation}
\rho (E)\equiv \frac{1}{2\pi }\sum_{m}A(m,E)=\frac{N_{LL}}{\left( 2\pi
\right) ^{2}\hbar }\sum_{n=0}^{\infty }\int 2\frac{d\epsilon (m)}{%
v_{z}(\epsilon (m),n)}A(\epsilon (m),E)  \label{DoS}
\end{equation}
The sum over LLs in (\ref{DoS}) can be again represented as a harmonic
series using the Poisson summation formula. Substituting (\ref{Rhohar}) into
(\ref{DoS}) we get 
\begin{eqnarray}
\rho (E) &=&\sum_{k=-\infty }^{\infty }(-1)^{k}\frac{N_{LL}}{\hbar \omega
_{c}}J_{0}\left( \frac{4\pi kt}{\hbar \omega _{c}}\right) \int \frac{%
d\epsilon (m)}{2\pi \hbar }A(\epsilon (m),E)\exp \left( \frac{2\pi
ik\epsilon (m)}{\hbar \omega _{c}}\right) =  \notag \\
&=&\frac{N_{LL}}{\hbar \omega _{c}}\sum_{k=-\infty }^{\infty
}(-1)^{k}J_{0}\left( \frac{4\pi kt}{\hbar \omega _{c}}\right) \exp \left( 
\frac{2\pi ik\,E}{\hbar \omega _{c}}\right) R_{D}(k,E)  \label{DoSHar}
\end{eqnarray}
From the formulas (\ref{ImS}, \ref{ImSTD} and \ref{DoSHar}), using $%
J_{0}(0)=1$ we get 
\begin{equation}
\left| \mbox{Im}\Sigma ^{R}(m,\epsilon )\right| =\pi k_{B}T_{D}\left(
1+2\sum_{k=1}^{\infty }(-1)^{k}J_{0}\left( \frac{4\pi kt}{\hbar \omega _{c}}%
\right) \cos \left( \frac{2\pi k\,\epsilon }{\hbar \omega _{c}}\right)
R_{D}(k,\epsilon )\right)  \label{ImS1}
\end{equation}
Together with (\ref{RD}) this is a nonlinear equation on $\mbox{Im}\Sigma
^{R}(m,\epsilon )$. We shall obtain explicit results in the next two
subsections only in the limit of strong harmonic damping where an expansion
over the harmonic damping factors in Eqs. (\ref{RD}) and (\ref{ImS1}) is
possible.

Substituting (\ref{ImS1}) into (\ref{Sigs}) we obtain the following
expression for the conductivity: 
\begin{eqnarray}
\sigma _{zz} &=&e^{2}N_{LL}\int d\epsilon \left( -n_{F}^{\prime }(\epsilon
)\right) \frac{\frac{2\left( td\right) ^{2}}{\hbar ^{2}\omega _{c}}\left( 1+%
\frac{\hbar \omega _{c}}{\pi t}\sum_{k=1}^{\infty }\frac{(-1)^{k}}{k}%
J_{1}\left( \frac{4\pi kt}{\hbar \omega _{c}}\right) \cos \left( \frac{2\pi
k\,\epsilon }{\hbar \omega _{c}}\right) R_{D}(k,\epsilon )\right) }{\pi
k_{B}T_{D}\left( 1+2\sum_{k=1}^{\infty }(-1)^{k}J_{0}\left( \frac{4\pi kt}{%
\hbar \omega _{c}}\right) \cos \left( \frac{2\pi k\,\epsilon }{\hbar \omega
_{c}}\right) R_{D}(k,\epsilon )\right) }+  \notag \\
&&+e^{2}N_{LL}\int d\epsilon \left( -n_{F}^{\prime }(\epsilon )\right) \frac{%
4td^{2}}{\hbar ^{2}\omega _{c}}\sum_{k=1}^{\infty }(-1)^{k}J_{1}\left( \frac{%
4\pi kt}{\hbar \omega _{c}}\right) \cos \left( \frac{2\pi k\,\epsilon }{%
\hbar \omega _{c}}\right) R_{D}(k,\epsilon )  \label{Sigd3}
\end{eqnarray}
Generally speaking, one should also take into account the oscillations of
the chemical potential, that are given by (\cite{Pavel2}, formula 5). The
formula (\ref{Sigd3}) is quite huge and without further simplification is
applicable only for the numerical calculation of the conductivity. In the
next two sections we shall analyze the limit of strong harmonic damping.

\section{The first order on damping factors; the phase shift of the beats.}

For simplicity we now assume the harmonic damping to be strong and neglect
all the terms that have additional powers of the damping factors: the Dingle
factor $R_{D}$ or the temperature smearing factor 
\begin{equation*}
R_{T}(k)\equiv \left( 2k\pi ^{2}k_{B}T/\hbar \omega _{c}\right) /\sinh
\left( 2k\pi ^{2}k_{B}T/\hbar \omega _{c}\right)
\end{equation*}
that will appear after integration over energy with the Fermi distribution
function. So, we rest only the first harmonic in the expression (\ref{Sigd3}%
) for the conductivity. This approximation is sufficient to obtain the phase
shift of the beating. In this approximation the Dingle factor $%
R_{D}(k,\epsilon )$ and the chemical potential can be considered as a
constant. The conductivity then becomes 
\begin{equation}
\sigma _{zz}=\frac{e^{2}N_{LL}}{2\pi }\int d\epsilon \left( -n_{F}^{\prime
}(\epsilon )\right) \frac{2td^{2}}{\hbar }\frac{1}{\pi k_{B}T_{D}}\frac{2t}{%
\hbar \omega _{c}}\times \frac{1}{1-2\cos \left( \frac{2\pi \epsilon }{\hbar
\omega _{c}}\right) J_{0}\left( \frac{4\pi t}{\hbar \omega _{c}}\right) R_{D}%
}-  \label{Sigd4}
\end{equation}
\begin{equation*}
-\frac{e^{2}N_{LL}}{2\pi }\int d\epsilon \left( -n_{F}^{\prime }(\epsilon
)\right) \frac{2td^{2}}{\hbar }\frac{1}{\pi k_{B}T_{D}}2\cos \left( \frac{%
2\pi \epsilon }{\hbar \omega _{c}}\right) J_{1}\left( \frac{4\pi t}{\hbar
\omega _{c}}\right) \left( 1+\frac{2\pi ^{2}k_{B}T_{D}}{\hbar \omega _{c}}%
\right) R_{D}
\end{equation*}
Since we consider the limit $R_{D}\ll 1$ the constant term is much larger
than the oscillating term in the denominator in the first line of Eq. (\ref
{Sigd4}). Hence, one can move up the small oscillating term using $%
1/(1+\alpha )\approx 1-\alpha ,\,\,\,\alpha \ll 1$. Then the integration
over $\epsilon $ can be easily performed. One gets 
\begin{equation}
\sigma _{zz}=\frac{e^{2}N_{LL}}{8\pi k_{B}T_{D}}\frac{\left( 4td\right) ^{2}%
}{\hbar ^{2}\omega _{c}}\left\{ 1+2\cos \left( \frac{2\pi \mu }{\hbar \omega
_{c}}\right) R_{T}R_{D}\left[ J_{0}\left( \frac{4\pi t}{\hbar \omega _{c}}%
\right) -\frac{\hbar \omega _{c}}{2\pi t}\left( 1+\frac{2\pi ^{2}k_{B}T_{D}}{%
\hbar \omega _{c}}\right) J_{1}\left( \frac{4\pi t}{\hbar \omega _{c}}%
\right) \right] \right\}  \label{Sig5}
\end{equation}
where $\mu $ is the chemical potential and the temperature smearing factor $%
R_{T}$ is given by the usual L-K expression: 
\begin{equation*}
R_{T}=\frac{2\pi ^{2}k_{B}T/\hbar \omega _{c}}{\sinh \left( 2\pi
^{2}k_{B}T/\hbar \omega _{c}\right) }
\end{equation*}

If the transfer integral is still large enough $4\pi t\gg \hbar \omega _{c}$
one can use the expansions of the Bessel function at large value of
argument: 
\begin{eqnarray}
J_{0}(x) &\approx &\sqrt{2/\pi x}\cos \left( x-\pi /4\right) \,,\,x\gg 1
\label{BFE} \\
J_{1}(x) &\approx &\sqrt{2/\pi x}\sin \left( x-\pi /4\right) \,,\,x\gg 1\,; 
\notag
\end{eqnarray}
Then the expression in square brackets of (\ref{Sig5}) simplifies and making
use of the standard trigonometric formulas one can write the oscillating
part of the conductivity as 
\begin{equation}
\tilde{\sigma}_{zz}=\frac{e^{2}N_{LL}}{8\pi k_{B}T_{D}}\frac{\left(
4td\right) ^{2}}{\hbar ^{2}\omega _{c}}\times 2\cos \left( \frac{2\pi \,\mu 
}{\hbar \omega _{c}}\right) R_{T}R_{D}\sqrt{\frac{\hbar \omega _{c}}{2\pi
^{2}t}}\sqrt{1+a^{2}}\cos \left( \frac{4\pi t}{\hbar \omega _{c}}-\frac{\pi 
}{4}+\phi \right)   \label{Sigd7}
\end{equation}
where 
\begin{equation}
a=\frac{\hbar \omega _{c}}{2\pi t}\left( 1+\frac{2\pi ^{2}k_{B}T_{D}}{\hbar
\omega _{c}}\right) \text{ \ \ and \ \ }\phi =\arctan \left( a\right) 
\label{aph2}
\end{equation}
One important difference of this formula from the usual Lifshitz-Kosevich
expression is the phase shift $\phi $ of the beats of the oscillations. This
phase shift can be experimentally observed by comparison with the dHvA
oscillations, that possess the phase offset consistent with the standard
Lifshitz-Kosevich formula and are proportional to 
\begin{equation}
\tilde{M}\sim \sin \left[ \frac{2\pi \mu }{\hbar \omega _{c}}\right]
R_{D}(T_{D})R_{T}(T)\cos (4\pi t/\hbar \omega _{c}-\pi /4)  \label{M}
\end{equation}

The above result concerning the phase of the beats of the SdH oscillations
has been recently qualitatively confirmed by a comparative experimental
study of the SdH and dHvA oscillations in an organic metal $\beta $%
-(BEDT-TTF)$_2$IBr$_2$ \cite{PhSh}.

The main limitation of the proposed analysis is that the magnetic field is
taken to be perpendicular to the conducting layers. A finite tilting angle $%
\theta $ of the magnetic field with respect to the normal to the conducting
planes may be approximately taken into account by rescaling the Landau level
separation: $\omega _{c}\rightarrow \omega _{c}\cos \theta $ and the warping
of the Fermi surface\cite{Yam} : $t(\theta )=t(0)\,J_{0}(k_{F}d\tan \theta )$%
, where $k_{F}$ is the in-plane Fermi momentum. But this is only a
semiclassical approximation based on the assumption that the FS remains the
same. Actually, the tilting of the magnetic field changes the dispersion
relation and a more profound study of the effect of the tilting magnetic
field on the transport properties is required. The quantum mechanical
calculation of the dispersion relation in tilted magnetic field in the first
order on the transfer integral\cite{Kur} gives the result close to that of
Yamaji\cite{Yam}; the correction is that the Bessel's function should be
replaced by the Laguerre polynomial\cite{Kur}: 
\begin{equation*}
J_{0}(k_{F}d\tan \theta )\rightarrow e^{-\left( A/2\right) ^{2}}\times
L_{n}\left( A^{2}/2\right) ,\,\,\,\text{where \ }A^{2}/2=d^{2}m\omega
_{c}\tan ^{2}\left( \theta \right) /2\hbar \text{ and }n=\mu /\hbar \omega
_{c}
\end{equation*}
But the use of the perturbation theory assumes that the transfer integral $t$
is much less than the Landau level separation $\hbar \omega _{c}$. Here we
consider the opposite case $2t>\hbar \omega _{c}$ where the beating can be
seen and some more profound study of the effect of the tilted magnetic field
is needed. For example, the effect of the tilting angle on the conductivity
should not be described just by this replacement near the zeros of $%
J_{0}(k_{F}d\tan \theta ).$ So, our analysis is applicable to a nonzer\qquad
o tilting angle only qualitatively.

Other errors may come from the approximate expression for the self energy (%
\ref{ImS}). The Born approximation in Q2D case works quite well, but maybe
other scattering mechanisms (especially for the calculation of the DoS)
should be taken into account. An accurate study of this problem may depend
on a particular type of a compound in hand.

The above analysis does not take into account the vertex corrections. In our
case this is valid because, according to the Ward identity, the vertex $\vec{%
\Gamma}(m,E)=\vec{p}+m\,\vec{\nabla}_{p}\Sigma ^{R}(m,E)$. Hence, if the
retarded self energy depends only on energy, the vertex corrections are
zero. The fact that $\Sigma ^{R}(m,\epsilon )$ is approximately a function
of only the energy $\epsilon $ is a consequence of the short-range (or
point-like) impurity potential. More precisely, if one takes a point-like
impurity potential and neglects all diagrams with the intersections of the
impurity lines in the self energy, then after averaging over randomly and
uniformly distributed impurity positions one obtains that $\Sigma
^{R}(m,\epsilon )=\Sigma ^{R}(\epsilon )$. The neglected graphs with the
intersections of the impurity lines describe the coherent scattering on two
impurities simultaneously. The contribution of such a scattering is small at
large enough interlayer transfer integral. In the three-dimensional case
without magnetic field the vertex corrections produce the factor $(1-\cos
\alpha )$ instead of $1$ in integrand for the transport scattering
relaxation time where $\alpha $ is the scattering angle. But if the
scattering probability is independent of the scattering angle (as for
point-like impurities) the additive $\cos \alpha $ vanishes after the
integration over angles and the vertex corrections vanish.

One should note that when the transfer integral is not much greater than the
cyclotron energy ($4\pi t/\hbar \omega _{c}\sim 1$) one cannot use the
expansions (\ref{BFE}) of the Bessel's functions and should apply the
formula (\ref{Sig5}) for the beatings of the conductivity oscillations. This
slightly changes the phase shift and makes the beatings not periodic. This
should be taken into account in the extraction of the transfer integral from
the beating frequency.

\section{The slow oscillations}

The slow oscillations come from the interference of the oscillations of
different quantities in (\ref{Sigd3}). Hence, one should consider the
entanglement of all oscillating quantities and find the non-oscillating term
in the lowest order on damping factors. For example, the Dingle factor
should no more be considered as constant because the entanglement of the
oscillations of the Dingle factor with other oscillating quantities produce
the term of the same order as the interference of the oscillations of the
mean square velocity and the scattering time. The errors in the self energy
now become more important, so, the subsequent analysis may not be
quantitatively very accurate. Nevertheless, it describes the main features
of the slow oscillations of the conductivity.

Up to the square of the Dingle factor in Eq. (\ref{Sigd3}) we get 
\begin{eqnarray}
\sigma _{zz} &=&\frac{e^{2}N_{LL}\left( 4td\right) ^{2}}{8\pi
k_{B}T_{D}\hbar ^{2}\omega _{c}}\int d\epsilon \left( -n_{F}^{\prime
}(\epsilon )\right) \times   \notag \\
&&\Biggl \{\left( 1-\frac{\hbar \omega _{c}}{\pi t}J_{1}\left( \frac{4\pi t}{%
\hbar \omega _{c}}\right) \cos \left( \frac{2\pi \,\epsilon }{\hbar \omega
_{c}}\right) R_{D}(\epsilon )\right) \times   \notag \\
&&\left[ 1+2J_{0}\left( \frac{4\pi t}{\hbar \omega _{c}}\right) \cos \left( 
\frac{2\pi \epsilon }{\hbar \omega _{c}}\right) R_{D}(\epsilon )+\left(
2J_{0}\left( \frac{4\pi t}{\hbar \omega _{c}}\right) \cos \left( \frac{2\pi
\epsilon }{\hbar \omega _{c}}\right) R_{D}(\epsilon )\right) ^{2}\right]  
\notag \\
&&-\frac{2\pi k_{B}T_{D}}{t}J_{1}\left( \frac{4\pi t}{\hbar \omega _{c}}%
\right) \cos \left( \frac{2\pi \,\epsilon }{\hbar \omega _{c}}\right)
R_{D}(\epsilon )\Biggr \}
\end{eqnarray}
where $R_{D}(\epsilon )$ is now given by (\ref{ImS}, \ref{ImSTD} and \ref{RD}%
) 
\begin{eqnarray}
R_{D}(\epsilon ) &\approx &\exp \left[ -\frac{2\pi ^{2}k_{B}T_{D}}{\hbar
\omega _{c}}\left( 1-2J_{0}\left( \frac{4\pi t}{\hbar \omega _{c}}\right)
\cos \left( \frac{2\pi \epsilon }{\hbar \omega _{c}}\right) R_{D}\right) %
\right] \approx   \notag \\
&\approx &R_{D}^{0}\,\left( 1+\frac{2\pi ^{2}k_{B}T_{D}}{\hbar \omega _{c}}%
2J_{0}\left( \frac{4\pi t}{\hbar \omega _{c}}\right) \cos \left( \frac{2\pi
\epsilon }{\hbar \omega _{c}}\right) R_{D}^{0}\right)   \label{RDa}
\end{eqnarray}
Combining these two expressions we get: 
\begin{equation}
\sigma _{zz}=\frac{e^{2}N_{LL}\left( 4td\right) ^{2}}{8\pi k_{B}T_{D}\hbar
^{2}\omega _{c}}\int d\epsilon \left( -n_{F}^{\prime }(\epsilon )\right)
\times \left( S_{1}+S_{2}\right)   \label{Sigd6}
\end{equation}
where 
\begin{equation}
S_{1}\equiv 1+2\cos \left( \frac{2\pi \epsilon }{\hbar \omega _{c}}\right) %
\left[ J_{0}\left( \frac{4\pi t}{\hbar \omega _{c}}\right) -\frac{\hbar
\omega _{c}}{2\pi t}\left( 1+\frac{2\pi ^{2}k_{B}T_{D}}{\hbar \omega _{c}}%
\right) J_{1}\left( \frac{4\pi t}{\hbar \omega _{c}}\right) \right] R_{D}
\label{So1}
\end{equation}
describes the constant term and the main term of the usual SdH oscillations
(they are the same as in (\ref{Sig5})), and 
\begin{equation}
S_{2}\equiv R_{D}^{2}2J_{0}\left( \frac{4\pi t}{\hbar \omega _{c}}\right) %
\left[ J_{0}\left( \frac{4\pi t}{\hbar \omega _{c}}\right) \left( 1+\frac{%
2\pi ^{2}k_{B}T_{D}}{\hbar \omega _{c}}\right) -\frac{\hbar \omega _{c}}{%
2\pi t}J_{1}\left( \frac{4\pi t}{\hbar \omega _{c}}\right) \left( 1+\frac{%
2\pi ^{2}k_{B}T_{D}}{\hbar \omega _{c}}+\left( \frac{2\pi ^{2}k_{B}T_{D}}{%
\hbar \omega _{c}}\right) ^{2}\right) \right]   \label{So2}
\end{equation}
describes the slow oscillations.

We shall consider the case $4t\gg \hbar \omega _{c}$ where our analysis
based on the self energy in the form (\ref{ImS}) has better accuracy and one
can use the large argument expansions (\ref{BFE}) of Bessel functions. The
formula (\ref{So2}) then simplifies: 
\begin{equation}
S_{2}\approx R_{D}^{2}\frac{\hbar \omega _{c}}{\pi ^{2}2t}\left( 1+\frac{%
2\pi ^{2}k_{B}T_{D}}{\hbar \omega _{c}}\right) \left[ 1+\sqrt{1+a_{S}^{2}}%
\cos \left( \frac{8\pi t}{\hbar \omega _{c}}-\frac{\pi }{2}+\phi _{S}\right) %
\right]  \label{SlOs}
\end{equation}
where 
\begin{equation}
a_{S}=\frac{\hbar \omega _{c}}{2\pi t}\left( 1+\frac{\left( 2\pi
^{2}k_{B}T_{D}/\hbar \omega _{c}\right) ^{2}}{1+2\pi ^{2}k_{B}T_{D}/\hbar
\omega _{c}}\right) \text{ \ \ and \ \ }\phi _{S}=\arctan \left( a_{S}\right)
\label{phs}
\end{equation}
We now have to consider the temperature and other smearing factors. But the
expression (\ref{SlOs}) does not depend on electron energy (unlike the
rapidly oscillating term (\ref{So1}) responsible for the SdH oscillations)
and, hence, all types of smearing coming from the averaging over the
electron energy do not affect the slow oscillations (they produce a factor $%
1 $). Hence, although the slow oscillations have a factor $R_{D}^{2}$
squared, they do not have the temperature damping factor $R_{T}$ entering
the SdH oscillations. Therefore, the slow oscillations may be much stronger
than the SdH oscillations. This was indeed observed at high enough
temperature\cite{MQORev,ibr2,i3}.

Now collecting the formulas (\ref{Sigd7}), (\ref{Sigd6}) and (\ref{SlOs})
and performing the integration over $\epsilon $ in (\ref{Sigd6}) one obtains
up to the second order on damping factors: 
\begin{eqnarray}
\sigma _{zz} &=&\frac{e^{2}N_{LL}}{8\pi k_{B}T_{D}}\frac{W^{2}d^{2}}{\hbar
^{2}\omega _{c}}\times \Bigg\{1+2\cos \left( \frac{2\pi \,\mu }{\hbar \omega
_{c}}\right) \sqrt{\frac{2\hbar \omega _{c}}{\pi ^{2}W}}\sqrt{1+a^{2}}\cos
\left( \frac{4\pi t}{\hbar \omega _{c}}-\frac{\pi }{4}+\phi \right)
R_{D}R_{T}+  \notag \\
&&+\frac{\hbar \omega _{c}}{\pi ^{2}2t}\left( 1+\frac{2\pi ^{2}k_{B}T_{D}}{%
\hbar \omega _{c}}\right) \left[ 1+\sqrt{1+a_{S}^{2}}\cos \left( \frac{8\pi t%
}{\hbar \omega _{c}}-\frac{\pi }{2}+\phi _{S}\right) \right] R_{D}^{2}\Bigg\}
\label{Sigf}
\end{eqnarray}

The formulas (\ref{Sigf}) and (\ref{phs}) make some predictions about the
slow oscillations that can be compared with the experimental data.

1). The frequency $F_{S}$ of the slow oscillations (of the conductivity as a
function of the inverse magnetic field) is two times larger than the beating
frequency. So, the distance between the two nearest nodes of the beatings
should be equal to the period of the slow oscillations.

2). The amplitude of the slow oscillations is temperature independent. Of
cause, if one consider the next terms in the expansion of the damping factor
the temperature dependence will appear (for example, from the oscillations
of the chemical potential). But these corrections are small. At $T\gtrsim
k_{B}T_{D}$ the amplitude of the slow oscillations is larger than the
amplitude of the SdH oscillations while at $T<k_{B}T_{D}$ the slow
oscillations are hardly distinguishable behind the SdH oscillations.

3). The phase shift of the slow oscillations is given by (\ref{SlOs}) and (%
\ref{phs}). It is equal to $\phi _{S}/2\approx \phi /2$ (the factor $1/2$ is
because the frequency of slow oscillations is two times greater than the
beating frequency). Hence the peaks of the slow oscillations must be between
the nearest peaks of the beats in the magnetization and conductivity. But
this phase is quite sensitive to the exact expression of the self energy.
Since we have used the approximate expression for $\mbox{Im}\Sigma
^{R}(m,\epsilon )$ (formula \ref{ImS}) the substantial error in the estimate
of $\phi _{S}$\ is possible. For example, the long range crystal
imperfections substantially damp the oscillations of the DoS that so that
the oscillations of the Dingle factor are much weaker than in formula (\ref
{RDa}). This long range electron scattering should be taken into account
when the $\mbox{Im}\Sigma ^{R}(m,\epsilon )$ is calculated.

\smallskip

The temperature independent harmonic damping factor comes not only from the
impurity scattering but also from the macroscopic inhomogeneities of a
sample. For example, the long-range background potential makes the electron
kinetic energy at the Fermi level slightly different over the sample that
produces the additional temperature independent harmonic damping. The
corresponding part of the Dingle temperature should not be included into (%
\ref{ImSTD}) and to the phase shifts (\ref{aph2}) and (\ref{phs}). This
smearing like the temperature smearing does not affect the phase shift and
the amplitude of the slow oscillations but causes an additional damping of
the SdH or dHvA oscillations. To make a quantitative estimate of the
amplitude of slow oscillations one can separate the impurity part of the
Dingle temperature from the comparison of the phase-shift and then apply the
formulas (\ref{Sigd6}), (\ref{So1}) and (\ref{SlOs}) without adjustable
parameters to compare the amplitudes of the oscillations. 

To summarize, we have calculated the main terms in the conductivity $\sigma
_{zz}$ in the layered compound in normal magnetic field. The obtained
formula (\ref{Sigf}) describes the slow oscillations and the phase shift of
the beatings. But the limitations of the calculation due to zero tilting
angle and the approximate form of the self energy may lead to some
quantitative deviations in the phase and the amplitude of the slow
oscillations.

We are thankful to A.M. Dyugaev and I. Vagner for the encouragement and
stimulating discussions. The work was supported by the EU ICN contract
HPRI-CT-1999-40013, and grants DFG-RFBR No. 436 RUS 113/592 and RFBR No.
00-02-17729a.

\appendix

\section{Appendix}

\subsection{Transformation of the sums over LLs to the sums over harmonics}

To transform the sums over LL number into the harmonic sums we shall apply
the Poisson summation formula \cite{ZW} 
\begin{equation}
\sum_{n=n_{0}}^{\infty }f(n)=\sum_{k=-\infty }^{\infty }\int_{a}^{\infty
}e^{2\pi ikn}f(n)\,dn\,,\,\text{\ \ where }a\in (n_{0}-1;n_{0})
\end{equation}
This formula is valid for arbitrary function $f(n).$ The sum in (\ref{Sigd2}%
) becomes 
\begin{eqnarray}
\sum_{n}\,\left| v_{z}(\epsilon ,n)\right| &=&\sum_{n=0}^{\infty }\frac{d} {%
\hbar }\sqrt{4t^{2}-\left( \epsilon -\hbar \omega _{c}\left( \,n+\frac{1}{2}
\right) \right) ^{2}}=  \notag \\
&=&\frac{d}{\hbar }\hbar \omega _{c}\sum_{k=-\infty }^{\infty
}\int_{0}^{\infty }dn\,\exp \left( 2\pi ik\left( n-\frac{1}{2}\right)
\right) \,\sqrt{\left( \frac{2t}{\hbar \omega _{c}}\right) ^{2}-\left( \frac{
\epsilon }{\hbar \omega _{c}}-\,n\right) ^{2}}  \notag \\
&=&\frac{d}{\hbar }\hbar \omega _{c}\sum_{k=-\infty }^{\infty }\left(
-1\right) ^{k}\exp \left( \frac{2\pi ik\epsilon }{\hbar \omega _{c}}\right)
\int_{-\infty }^{\infty }dx\,\exp \left( 2\pi ikx\right) \,\sqrt{\left( 
\frac{2t}{\hbar \omega _{c}}\right) ^{2}-x^{2}}  \notag \\
&=&\sum_{k=-\infty }^{\infty }\frac{dt}{\hbar k}\left( -1\right) ^{k}\exp
\left( \frac{2\pi ik\epsilon }{\hbar \omega _{c}}\right) J_{1}\left( \frac{
4\pi kt}{\hbar \omega _{c}}\right)  \label{Sv}
\end{eqnarray}
In an analogous way one can also transform the sum 
\begin{equation}
\frac{\pi }{N_{LL}}\rho _{0}(\epsilon )=\sum_{n=0}^{\infty }\frac{1}{\sqrt{
4t^{2}-\left( \epsilon -\hbar \omega _{c}\left( \,n+\frac{1}{2}\right)
\right) ^{2}}}=  \label{rhon}
\end{equation}
\begin{eqnarray}
&=&\frac{1}{\hbar \omega _{c}}\sum_{k=-\infty }^{\infty }\int_{0}^{\infty } 
\frac{dn\,\exp \left( 2\pi ik\left( n-\frac{1}{2}\right) \right) }{\,\sqrt{%
\left( \frac{2t}{\hbar \omega _{c}}\right) ^{2}-\left( \frac{\epsilon }{
\hbar \omega _{c}}-\,n\right) ^{2}}}=  \notag \\
&=&\frac{1}{\hbar \omega _{c}}\sum_{k=-\infty }^{\infty }\left( -1\right)
^{k}\exp \left( \frac{2\pi ik\epsilon }{\hbar \omega _{c}}\right)
\int_{-\infty }^{\infty }\frac{dx\,\exp \left( 2\pi ikx\right) \,}{\sqrt{%
\left( \frac{2t}{\hbar \omega _{c}}\right) ^{2}-x^{2}}}=  \notag \\
&=&\frac{\pi }{\hbar \omega _{c}}\sum_{k=-\infty }^{\infty }\left( -1\right)
^{k}\exp \left( \frac{2\pi ik\epsilon }{\hbar \omega _{c}}\right)
J_{0}\left( \frac{4\pi kt}{\hbar \omega _{c}}\right)  \label{Rhohar}
\end{eqnarray}

\end{document}